\documentclass[manuscript]{aastex63}
\usepackage{graphicx}
\usepackage{subfigure}
\usepackage{CJK}
\usepackage{amsmath,bm}
\usepackage{comment}
\usepackage{appendix}
\usepackage{lineno}

\received{----}
\revised{\today}
\accepted{----}
\submitjournal{APJ}

\begin{document}
\begin{CJK*}{UTF8}{gbsn}

\title{Propagation and transmission of Jupiter's internal waves}

\author[0009-0000-8067-4491]{Yuru Xu (续育茹)}
\affiliation{IFAA, School of Physics and Astronomy, Beijing Normal University, Beijing 100875, China}

\author[0000-0002-3641-6732]{Xing Wei (魏星)}
\affiliation{IFAA, School of Physics and Astronomy, Beijing Normal University, Beijing 100875, China}

\correspondingauthor{Xing Wei}
\email{xingwei@bnu.edu.cn}

\begin{abstract}
Observations from the Juno spacecraft show that Jupiter has a large dilute core rather than a compact core. To investigate the effects of different core structures on wave propagation and transmission in Jupiter's interior, we consider three models: (1) an isentropic sphere, (2) an isentropic envelope with a rigid core, and (3) an isentropic envelope with a dilute core. We study the propagation and transmission of p modes (sound waves), g modes (gravity waves), r modes (inertial waves), and GIWs (gravito-inertial waves) by solving the linear equations of a compressible, self-gravitating, uniformly rotating polytropic model, fully taking into account the the effects of Coriolis force but neglecting centrifugal flattening. Our results show that energy flux is primarily carried by fast waves with higher frequencies whereas kinetic energy by slow waves with lower frequencies. Rotation has a greater effect on non-axisymmetric modes than on axisymmetric ones. In model 2, rigid core facilitates propagation of r modes. In model 3, rotation enhances the transmission of GIWs across the interface between the dilute core and the isentropic envelope, particularly at high latitudes. This suggests that Jupiter's internal structure may be inferred by detecting the oscillation signals in its polar regions.
\end{abstract}


\section{Introduction} \label{sec:intro}

Understanding the internal structure of stars and planets is a crucial topic in astrophysics and planetary science. As the most massive planet in the solar system, Jupiter has long been a focus of theoretical and observational studies. Studying Jupiter's internal structure provides key insights into the origin of planetary systems, the behavior of materials under high pressure and temperature, and the diverse properties of giant planets \citep{2004jpsm.book...35G,2023ASPC..534..947G,helled2022revelations,stevenson2022mixing,miguel2023interior,helled2023nature}. Historically, the internal structure of giant planets has been modeled using a three layer model: a small compact core (rocky or icy) beneath a convective envelope of metallic hydrogen and helium, surrounded by an outer atmospheric envelope of molecular hydrogen and helium \citep{1982AREPS..10..257S, 2005AREPS..33..493G, 2010SSRv..152..423F}. Each layer is typically assumed to be chemically homogeneous, with heavy elements concentrated in the core. While the traditional three-layer model provides a useful starting point, uncertainties remain regarding the size of each region and the transitions between them. Moreover, these models fail to explain all observations. Precise measurements of Jupiter's gravitational field from the Juno mission \citep{2017Sci...356..821B} have introduced new challenges to conventional structure models. In these updated models, Jupiter's core is no longer considered as a compact, heavy-element region with a density discontinuity at the core-envelope boundary \citep{2017GeoRL..44.4649W,2017ApJ...840L...4H,2019ApJ...872..100D,2022PSJ.....3..185M,2022PSJ.....3...89I,2023A&A...672A..33H,2024Icar..41115955M}. Instead, the mass of Jupiter is distributed throughout the interior of the planet, existing as a large dilute core that extends into the region previously thought to be convective.

Wave propagation allows us to \textquotesingle see\textquotesingle \, the internal structures of celestial bodies as in seismology, helio- and asteroseismology \citep{RevModPhys.74.1073, gilliland2010kepler, aerts2010asteroseismology, chaplin2013asteroseismology, 2014AGUFM.P21E..03F,mankovich2019cassini,mankovich2021diffuse,aerts2021probing,mankovich2023saturn}. Additionally, internal waves in a star or planet can transport momentum, angular momentum and energy while propagating, so that wave propagation is important for studying the internal structure and evolution of stars and planets \citep{1979nos..book.....U,2005ApJ...635..674W,2005ApJ...635..688W,2008Natur.456..770T,Goodman2008DYNAMICALTI,2013ApJ...772...21R}. \citet{1999ApJ...521..764L} studied inertial modes in a sphere with an arbitrary polytropic density profile. \citet{2005ApJ...635..688W} found that resonantly excited inertial modes significantly influence the tidal dissipation of a coreless Jupiter. \citet{Goodman2008DYNAMICALTI} argued that a rigid core should be incorporated into the model. In their discussion, inertial waves were assumed to be fully reflected at the rigid core's surface. Recent studies have focused on the dilute core model of Jupiter's dynamical tide \citep{lai2021jupiter,2022PSJ.....3...89I,2023MNRAS.521.5991D,lin2023dynamical}. However, if the rigid core is replaced by a dilute core, inertial waves are expected to be partially reflected at the interface. This can be regarded as a problem of wave propagation and transmission at the interface between a stably stratified layer (dilute core is treated as a stably stratified layer) and a convective region. The stably stratified layer that supports g modes (gravity waves) restored by buoyancy forces. If the information of g modes can be detected on Jupiter's surface, then it could be possible to detect Jupiter's deep interior. The purpose of this paper is to investigate the effects of different core structures on wave propagation and transmission in Jupiter's interior.

The transmission of internal waves through a density staircase in a local Cartesian model was investigated by \citet{PhysRevFluids.1.050509}. They used the \textquotesingle traditional approximation\textquotesingle \, to account for rotation, which assumes that the buoyancy force dominates over the Coriolis acceleration along the direction of stratification, thereby disallowing inertial waves. Density staircases can result from semiconvective layers or layered convection \citep{2012ApJ...750...61M,2016ApJ...817...54M,2022PhRvF...7l4501F,2024ApJ...975L...1F}. \citet{2017A&A...605A.117A} studied the transmission of gravity and inertial waves in a similar Cartesian geometry. Their results showed that a density staircase strongly affects wave transmission, with gravito-inertial waves being more readily transmitted when their wavelengths are large compared to the step size. Subsequently, \citet{2020MNRAS.493.5788P} examined the propagation and transmission of internal waves using a simplified global Boussinesq planetary model without rotation. Furthermore, \citet{2020ApJ...890...20W,2020ApJ...899...88W} studied the reflection and transmission of an incident wave at the stratification-convection boundary in a local Cartesian (f-plane) model, deriving an analytical expression for the transmission ratio and demonstrated that rotation significantly enhances wave transmission. On Wei's work, \citet{cai2021inertial} considered two different stratification structures in the stable layer: a uniform stratified layer, or a continuously varying stratified layer. Efficiencies of wave transmissions were estimated and compared.

 In this paper, we model Jupiter with a rotating polytrope of index $n=1$, which serves as a suitable leading-order approximation of its internal structure \citep{2004ApJ...610..477O,2005ApJ...635..674W,2005ApJ...635..688W,2020AREPS..48..465S,lai2021jupiter,2021PSJ.....2...69I,2022PSJ.....3...89I,2022PSJ.....3...11I,2022ApJ...925..124D,2023MNRAS.521.5991D,lin2023dynamical}. This global (spherical) and fully compressible model introduces numerical challenges due to the presence of p modes, i.e., sound waves supported by pressure forces. Moreover, we are concerned with rotation to explore its influence on wave propagation and transmission. Notably, because of Coriolis force our problem becomes a two-dimensional eigenvalue calculation \citep{1999JFM...398..271D,2006A&A...455..621R}. The structure of this paper is as follows. In Section \ref{sec:modes in jupiter}, we outline our research model and numerical methods. In Section \ref{sec:results}, we consider three models (full sphere, compact rigid core and dilute core) of Jupiter's internal structure, focusing on the effects of different core structures on the propagation and transmission of internal waves, as well as the influence of rotation. Finally, in Section \ref{sec:conclusions}, we summarize the conclusions, briefly discuss the implications of our results and suggest the future study.

\section{equations and models} \label{sec:modes in jupiter}
In this section, we describe the method to calculate Jupiter's internal structure and the normal mode oscillations. 

\subsection{Equations}
The rotational effect is measured by $\epsilon = \Omega/\omega_{r}$, i.e., the ratio of rotation rate $\Omega$ to dynamical frequency $\omega_{r}=\left ( GM/R^{3} \right ) ^{1/2}$ where $G$ is the gravitational constant and $M$ and $R$ are respectively planetary mass and radius. Jupiter's $\epsilon=0.288$ is large. The ratio of Coriolis force to self-gravity is at the order $O(\epsilon)$ and that of centrifugal force to self-gravity is at $O(\epsilon^2)$. Although Jupiter's centrifugal force induces an equatorial deformation, we neglect this second-order centrifugal effect compared to the first-order Coriolis effect. Therefore, we assume a spherically symmetric equilibrium of polytropic fluid, and in the perturbation the full Coriolis effect is taken into account.
 
The equilibrium state is modeled as a self-gravitating, uniformly rotating polytrope, described by the following equations \citep{aerts2010asteroseismology}:
\begin{equation} 
    P_{0}=K\rho_{0}^{1+1/n},
\end{equation}
\begin{equation} 
    \boldsymbol{\nabla}P_{0}=-\rho_{0}\boldsymbol{\nabla}\Phi_{0},
\end{equation}
\begin{equation} 
    \nabla^{2}\Phi_{0}=4\pi G\rho_{0},
\end{equation}
where $P$ is the pressure, $\rho$ the density, $K$ the polytropic constant, $n$ the polytropic index, and $\Phi$ the gravitational potential. The subscript ‘${0}$’ denotes the equilibrium state. In our calculations, we normalize equations with Jupiter's mass and radius, such that the time unit is the dynamical time scale $1/\omega_{r}$.

We calculate adiabatic and inviscid oscillation modes by Eulerian perturbation to the equilibrium state. The perturbation is assumed to have a time dependence of the form $\propto e^{-i\omega t}$, where $\omega$ represents the mode frequency in the rotating frame. The linearized fluid dynamics equations in spherical coordinates $\left(r, \theta, \phi\right)$ are given by the following set of equations, as described in \citet{lin2023dynamical} (without tidal and viscous forces)
\begin{equation} \label{eq:mom}
    -i\omega \bm u^{\prime} =-\frac{1}{\rho_{0}}\boldsymbol{\nabla}\left(\rho_{0}h^{\prime}\right)+2\boldsymbol{u^{\prime}\times \Omega}+\frac{\boldsymbol{g}_{0}}{4\pi G\rho_{0}}\nabla^{2}\Phi^{\prime}-\boldsymbol{\nabla}\Phi^{\prime},    
\end{equation}
\begin{equation} 
    -i\omega h^{\prime} =-c_{0}^{2}\left[\frac{N_{0}^{2}}{g_{0}}u_{r}^{\prime}+\frac{1}{\rho_{0}}\boldsymbol{\nabla}\cdot \left ( \rho_{0}\boldsymbol{u^{\prime}}\right )\right],
\end{equation}
\begin{equation} \label{eq:possion2}
    -i\omega\nabla^{2}\Phi^{\prime}=-4\pi G\boldsymbol{\nabla}\cdot \left ( \rho_{0}\boldsymbol{u^{\prime}}\right ),    
\end{equation}
where $\boldsymbol{u}'$ is the velocity, $h=p^{\prime}/\rho_{0}$ the enthalpy perturbation, $u_{r}'$ the radial velocity, $c_{0}^{2}$ the square of the adiabatic sound speed, $N_{0}^{2}$ the square of buoyancy frequency and $\boldsymbol{g}_{0}$ the gravitational acceleration. The superscript prime denotes Eulerian perturbations. The coefficients $\boldsymbol{g}_{0}$, $c_{0}^{2}$, $N_{0}^{2}$ are determined by the equilibrium
\begin{equation} 
     \boldsymbol{g}_{0}=-\boldsymbol{\nabla}\Phi_{0},
\end{equation}
\begin{equation} 
	c_{0}^{2}=\Gamma P_{0}/\rho_{0}, 
\end{equation}
\begin{equation} 
	N_{0}^{2}=\boldsymbol{g}_{0}\cdot\left ( -\frac{1 }{\Gamma} \frac{\boldsymbol{\nabla }P_{0}}{P_{0}}+\frac{\boldsymbol{\nabla }\rho_{0}}{\rho_{0}}  \right),
\end{equation}
where $\Gamma=\left(\partial \,ln\,P/\partial \,ln\,\rho\right)_{ad}$ is the adiabatic index. The convective stability of a convective or stably stratified layer can be measured by the square of buoyancy frequency $N_{0}^2$. A convective layer has $N_{0}^2\leq 0$ ($N_{0}^2=0$ indicates neutral stability), while a stably stratified layer has $N_{0}^2>0$. Since convection efficiently transports energy, the thermal structure in the convective layer is nearly adiabatic, resulting in $N_{0}^2\approx 0$.

In order to solve Eqs.(\ref{eq:mom}-\ref{eq:possion2}), The variables $\boldsymbol{u}^{\prime}$, $h^{\prime}$ and $\Phi^{\prime}$ are expanded as a sum over the spherical harmonics:
\begin{equation} 
\boldsymbol{u}^{\prime}=\sum_{l=m}^{L}u_{l}^{m}\left(r\right)\boldsymbol{R}_{l}^{m}+\sum_{l=m}^{L}v_{l}^{m}\left(r\right)\boldsymbol{S}_{l}^{m}+\sum_{l=m}^{L}w_{l}^{m}\left(r\right)\boldsymbol{T}_{l}^{m},
\label{eq:u}
\end{equation}
\begin{equation} 
	h^{\prime}=\sum_{l=m}^{L}h_{l}^{m}\left(r\right)Y_{l}^{m}\left(\theta,\phi\right),
 \label{eq:h}
\end{equation}
\begin{equation} 
	\Phi^{\prime}=\sum_{l=m}^{L}\Phi_{l}^{m}\left(r\right)Y_{l}^{m}\left(\theta,\phi\right),
    \label{eq:phi}
\end{equation}
where $Y_{l}^{m}$ is the spherical harmonic functions of degree $l$ and azimuthal order $m$, while $L$ denotes the truncation degree applied in the numerical calculations. Given that the equilibrium state is axisymmetric, the variable $\phi$ does not couple with the other two variables, $r$ and $\theta$. As a result, there is no summation over the order $m$ in these expressions. But due to the presence of Coriolis force, the radial and angular components of the perturbation variables are coupled and cannot be separated, so that summation is taken over the degree $l$. $\boldsymbol{R}_{l}^{m}$, $\boldsymbol{S}_{l}^{m}$ and $\boldsymbol{T}_{l}^{m}$ are vector spherical harmonics:
\begin{equation} 
	\boldsymbol{R}_{l}^{m}=Y_{l}^{m}\left(\theta,\phi\right)\Tilde{\boldsymbol{r}},\quad \boldsymbol{S}_{l}^{m}=r\boldsymbol{\nabla}Y_{l}^{m}\left(\theta,\phi\right),\quad
    \boldsymbol{T}_{l}^{m}=r\boldsymbol{\nabla}\times \boldsymbol{R}_{l}^{m}.
\end{equation}
By Substituting the expressions Eqs.(\ref{eq:u}-\ref{eq:phi}) into Eqs.(\ref{eq:mom}-\ref{eq:possion2}) and projecting the equations onto spherical harmonics, we derive a series of ordinary differential equations (ODEs) involving $u_{l}^{m}\left(r\right)$, $v_{l}^{m}\left(r\right)$,$w_{l}^{m}\left(r\right)$, $h_{l}^{m}\left(r\right)$, and $\phi_{l}^{m}\left(r\right)$. We use a resolution of Chebyshev collocation on 100 Gauss-Lobatto nodes in the radial direction and set the truncation degree $L$ to range from 60 to 80.

To solve the ODEs we need to impose the boundary conditions. The Lagrangian perturbation of pressure vanishes at surface $\delta P=P_{1}+u_{r}/\left(-i\omega\right)\nabla P_{0}=0$ \citep{Goodman2008DYNAMICALTI,2021PSJ.....2..198D}. This free-surface boundary condition will induce the surface gravity wave (f mode) whose frequency is comparable to that of sound wave due to the hydrostatic balance at surface. In terms of $h^{\prime}$ and $u_{r}^{\prime}$, the boundary condition can be rewritten with Eqs.(\ref{eq:u}-\ref{eq:h}) as
\begin{equation} \label{eq:pressure_outer}
	-i\omega h_{l}^{m}=g_{0}u_{l}^{m} \quad at \quad r=R.    
\end{equation}
The zero radial velocity $u_{r}=0$ on the rigid inner boundary is given as \citep{Goodman2008DYNAMICALTI}
\begin{equation} \label{eq:ur_inner}
	u_{l}^{m}=0 \quad at\quad r=R_{i}.    
\end{equation}
The boundary conditions of the gravitational perturbations at both inner and outer boundaries are given as 
\begin{equation} 
	r\frac{d\Phi_{l}^{m}}{dr}+\left(l+1\right)\Phi_{l}^{m}=0 \quad at \quad r=R,
\end{equation}
\begin{equation} 
	r\frac{d\Phi_{l}^{m}}{dr}-l\Phi_{l}^{m}=0 \quad at\quad r=R_{i}.
\end{equation}
Now the system reduces to a generalized eigenvalue problem $\mathcal{A}X=-i\omega\mathcal{B}X$ where $\mathcal{A}$ and $\mathcal{B}$ are coefficient matrices. The eigensolutions ($\omega$,$X$) correspond to the frequencies and spatial distributions of the oscillations. We use a standard direct solver to compute all eigensolutions for each case.

\subsection{Interior models}
Our understanding of Jupiter's interior has been greatly improved by Juno observations \citep{2017GeoRL..44.4649W,2017Sci...356..821B,2019ApJ...872..100D,2020AREPS..48..465S,nettelmann2021theory,2022PSJ.....3..185M,2024Icar..41115955M}, though some uncertainties still remain. In this paper，we do not aim to construct a fully realistic model of Jupiter's interior. Instead, we focus on the effects of different core structures on wave propagation and transmission. In all the three models (1) full sphere, (2) rigid core and (3) dilute core, the equilibrium profiles $\rho_{0}\left(r\right)$, $g_{0}\left(r\right)$ are given by $n=1$. The $n=1$ polytrope closely approximates the equation of state (EOS) of H-He, the elements that dominate the composition of gas giant planets \citep{2020AREPS..48..465S}. The adiabatic index $\Gamma$, which is determined by the chemical compositions, is constant in Models 1 and 2 but varies in Model 3  \citep{lai2021jupiter,lin2023dynamical,2023MNRAS.521.5991D}.

\begin{figure}
    \subfigure[]{
 \includegraphics[width=0.45\textwidth,height=6cm]{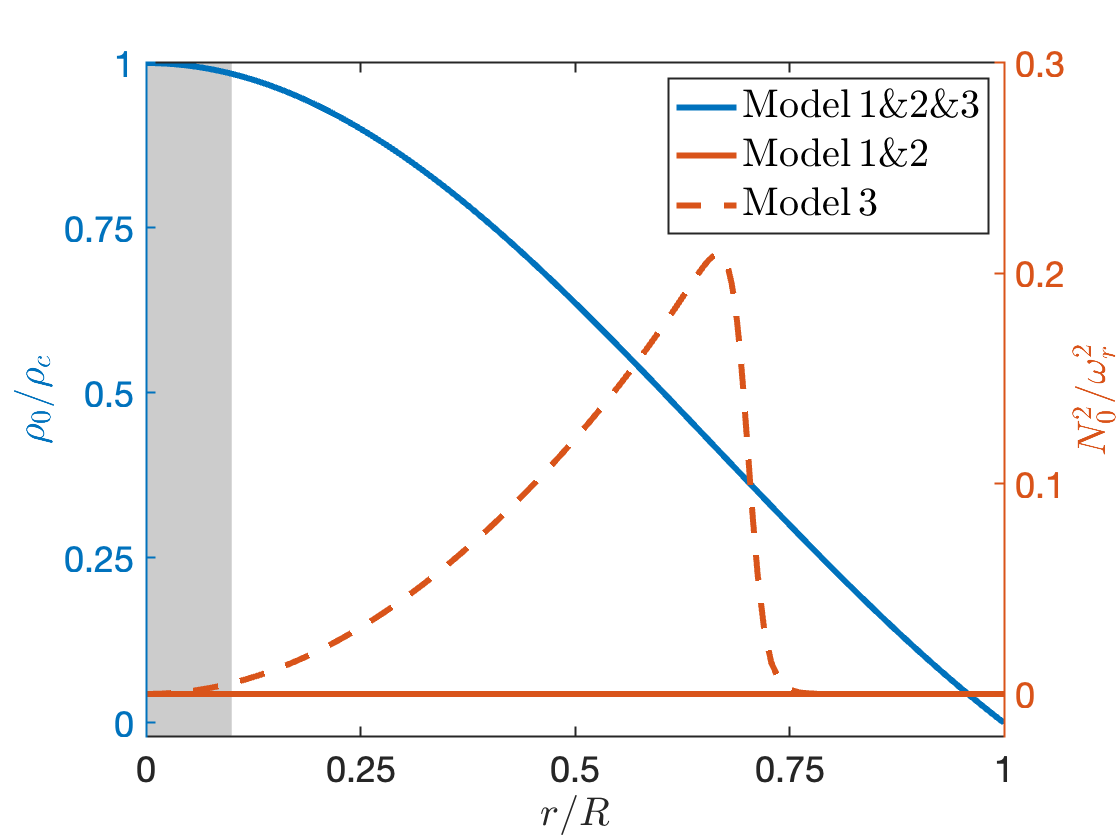}
    \label{fig:bg1}}
    \subfigure[]{    \includegraphics[width=0.45\textwidth,height=6cm]{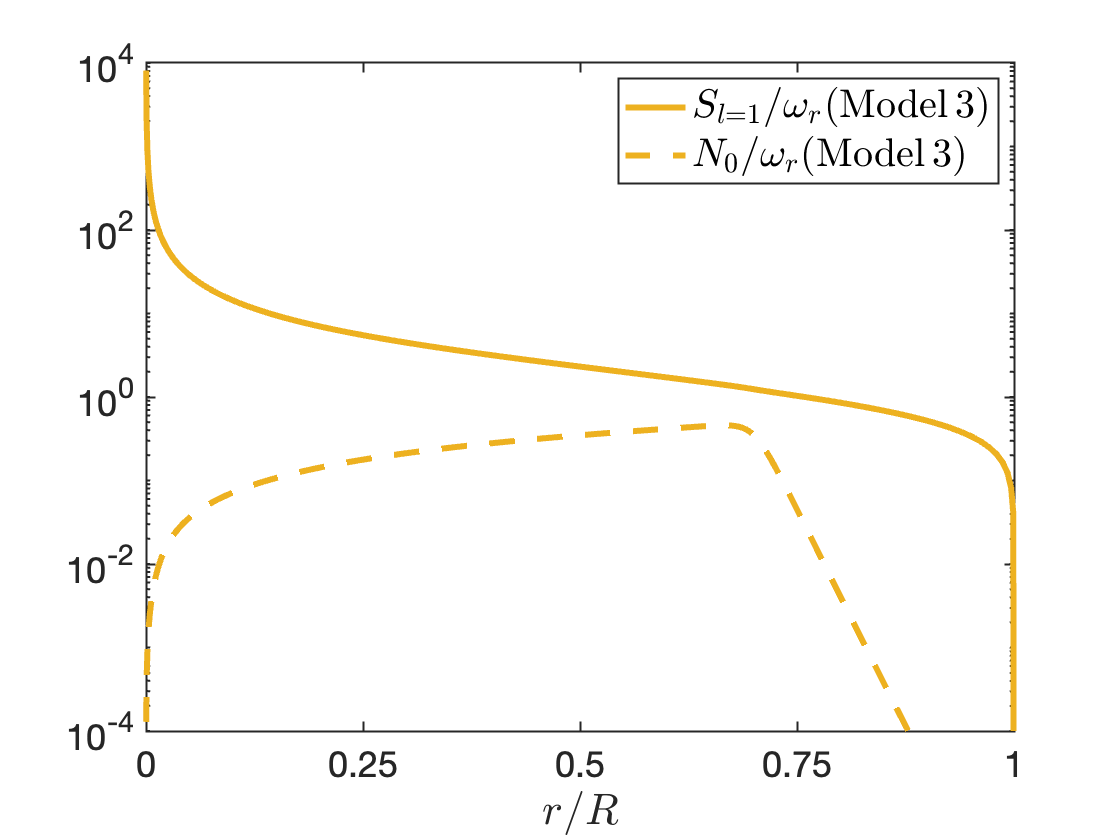}
    \label{fig:bg2}}
    \caption{(a) Three models of Jupiter's interior. The blue solid line represents the normalized density as a function of the normalized radius. The red solid line indicates the square of the normalized buoyancy frequency for model 1 (full sphere model) and model 2 (compact rigid core model), while the red dashed line shows the square of the normalized buoyancy frequency for model 3 (dilute core model). The gray shadow denotes solid regions. (b) Model 3 of dilute core. The yellow solid line shows the normalized sound wave frequency with $l=1$ and the yellow dashed line the normalized buoyancy frequency. \label{fig:bg}}
\end{figure}

Figure \ref{fig:bg1} shows the normalized density $\rho_{0}/\rho_{c}$ and the squre of buoyancy frequency $N_{0}^{2}/\omega_{r}$ profiles as a function of the radius for the three nominal models considered in this paper. Model 1 is a full isentropic polytrope with $\Gamma=1+1/n=2$ and $N_{0}^{2}/\omega_{r}=0$ throughout the entire fluid sphere, serving as a reference for the other two models. Model 2 features a small rigid core with a radius of $0.1R$ (the grey shadow indicates solid region) and an isentropic fluid envelope, i.e., $\Gamma=2$ and $N_{0}^{2}/\omega_{r}=0$ in the fluid region. The recent interior model \citep{2024Icar..41115955M} and evolutionary model \citep{2024ApJ...971..104S,2025ApJ...979..243A} for Jupiter's gravity harmonics $J_{2}$ and $J_{4}$ suggest that if Jupiter has a compact core then its mass should be less than approximately 3 Earth masses and its radius is around 0.1 Jupiter radii. In this model we focus only on the effect of the presence of a rigid core on wave propagation and transmission but do not consider the realistic model. The dependence of wave propagation and transmission on the size of core will be explored in future work. Model 3, as assumed by \citet{lin2023dynamical} and \citet{2023MNRAS.521.5991D}, consists of an extended, dilute core with a radius of 0.7R and an isentropic envelope. The dilute core is treated as a stably stratified fluid layer with $\Gamma>2$ expressed by \citep{lai2021jupiter}
\begin{equation}
\Gamma=2+\frac{0.1}{\left(1+e^{100(r/R-0.7)}\right)\left(1+e^{-100r/R}\right)}.
\end{equation}
In addition to adjusting the adiabatic index ($\Gamma>2$), the stable stratification can be achieved in the diluted core by adjusting the density profile while keeping the adiabatic index $\Gamma=2$ \citep{2022PSJ.....3...89I}. It should be noted that these profiles (density and buoyancy frequency) in our three models are employed only to capture the relevant wave dynamics but not the realistic internal models.

Figure \ref{fig:bg2} shows the normalized sound wave frequency $S_l/\omega_{r}=\sqrt{l(l+1)}c_0/r$ (of the fundamental $l=1$) and buoyancy frequency in Model 3 of dilute core. We can see that the two frequencies are comparable to each other, which implies that the p and g modes are already mixed.

\section{results} \label{sec:results}

In this section, we show the results of the oscillation modes in the three models. To investigate the effect of the Coriolis force we test various rotation rates. We analyze the perturbed radial energy flux and total kinetic energy carried by these waves. The perturbed radial energy flux $F$ and kinetic energy $K$ are given by (see the derivation details in Appendix \ref{appendix})
\begin{equation}
	F=\left(1+n\right)p^{\prime}{u_r}^{\prime}+\Phi_{0}\rho^{\prime}{u_r}^{\prime},
\end{equation}
\begin{equation}
    K=\rho_0 u'^2.
\end{equation}

\subsection{Full sphere model}
\begin{figure}
    \centering
        \includegraphics[width=0.9\textwidth]{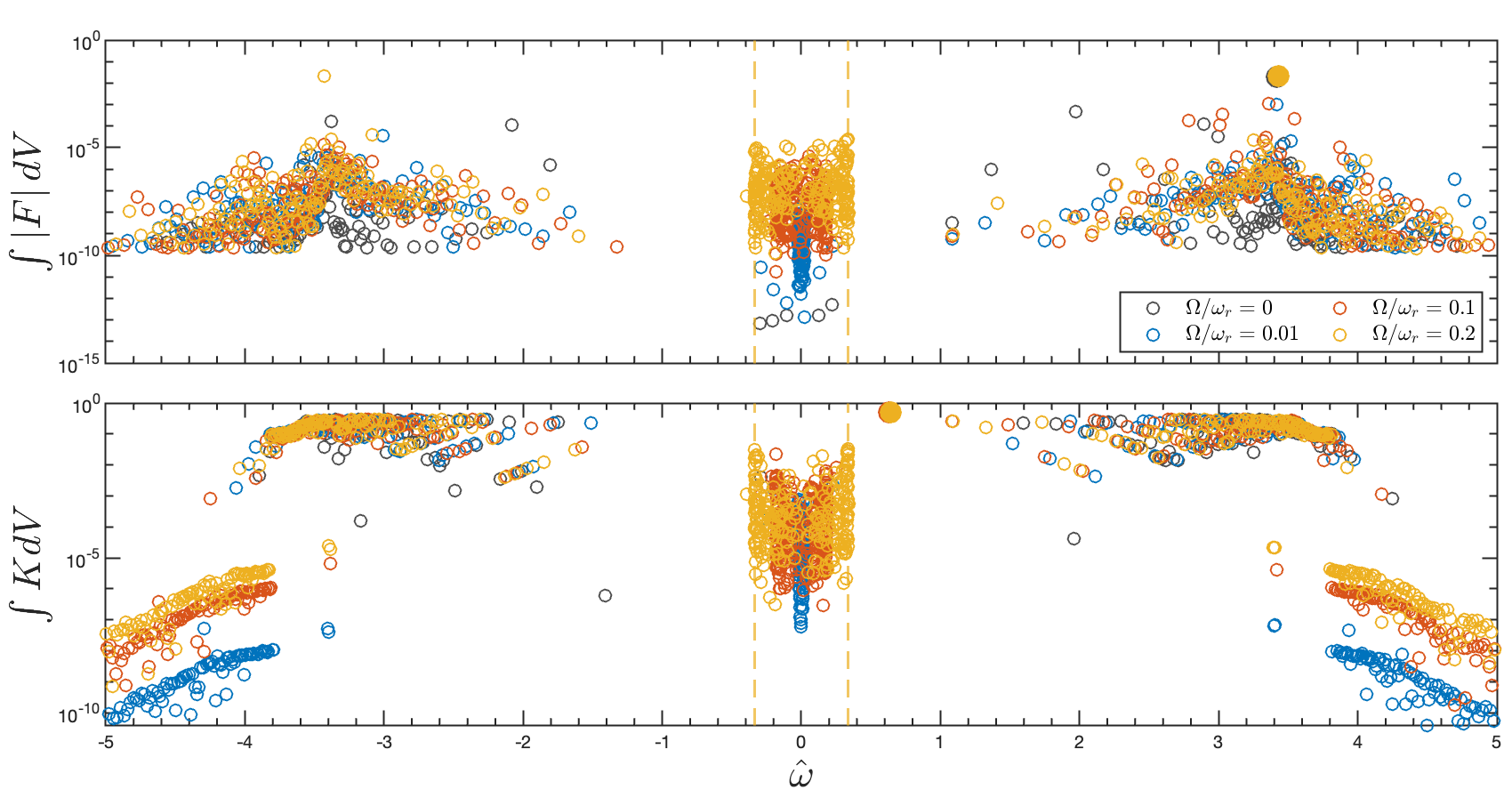}
    \caption{Model 1 with $m=0$. Radial energy flux and kinetic energy as a function of wave frequency $\hat{\omega}$ ($\hat{\omega}=sign(\omega)*\ln\left(1+\left|\omega/\omega_{r}\right|\right)$). Top panel shows radial energy flux and bottom panel shows the kinetic energy. Different colors denote different rotation rates: $\Omega/\omega_{r}=0$ (grey), $\Omega/\omega_{r}=0.01$ (blue), $\Omega/\omega_{r}=0.1$ (red), and $\Omega/\omega_{r}=0.2$ (yellow). Solid dots represent $\hat{\omega}$ corresponding to max$\int \left|F\right|dV$ or max$\int KdV$($dV=r^2\sin\theta drd\theta d\phi$). For $\Omega/\omega_{r}=0.2$, the frequency of r modes lies between the two yellow dashed lines, while the region outside corresponds to p modes.}
    \label{fig:n1m0}
\end{figure}

We begin with the wave propagation of isentropic fluid in a full sphere. In the absence of rotation, only p modes (sound waves) are excited by pressure \citep{1979nos..book.....U,aerts2010asteroseismology}. In the presence of rotation, r modes (inertial waves) can be excited by Coriolis forces within the frequency range $0 < \left|\omega\right| < 2\Omega$ \citep{2004ApJ...610..477O,2005ApJ...635..674W,Goodman2008DYNAMICALTI,2010MNRAS.407.1631P,2014ARA&A..52..171O}. Fig.\ref{fig:n1m0} shows the absolute value of the perturbed radial energy flux $\left|F\right|$ and kinetic energy $K$ in the full sphere with the azimuthal wavenumber $m=0$. To more clearly see the wave frequency near zero, we apply a zoom-in transformation $\hat{\omega}=sgn\left(\omega\right)*\ln\left( 1+\left|\omega/\omega_{r}\right|\right)$. For $\Omega/\omega_{r}=0.2$, r modes lie between the two yellow dashed lines, while the region outside corresponds to p modes. Solid dots represent the wave frequency $\hat{\omega}$ corresponding to max $\int\left|F\right|dV$ or max $\int K dV$ ($dV=r^2\sin\theta dr d\theta d\phi$). The result indicates that both radial energy flux and kinetic energy are dominated by p modes, and rotation has little effect on wave frequency of max $\int\left|F\right|dV$ or max $\int K dV$. Moreover, the wave associated with $max\int\left|F\right|dV$ corresponds to fast p mode with higher frequency, while that associated with $max\int KdV$ to slow p mode with lower frequency. We infer that higher frequency waves with faster propagation tend to transport a significant portion of energy outward, whereas lower frequency waves with slower propagation tend to retain energy.

\begin{figure}
    \centering
      \includegraphics[width=0.9\textwidth]{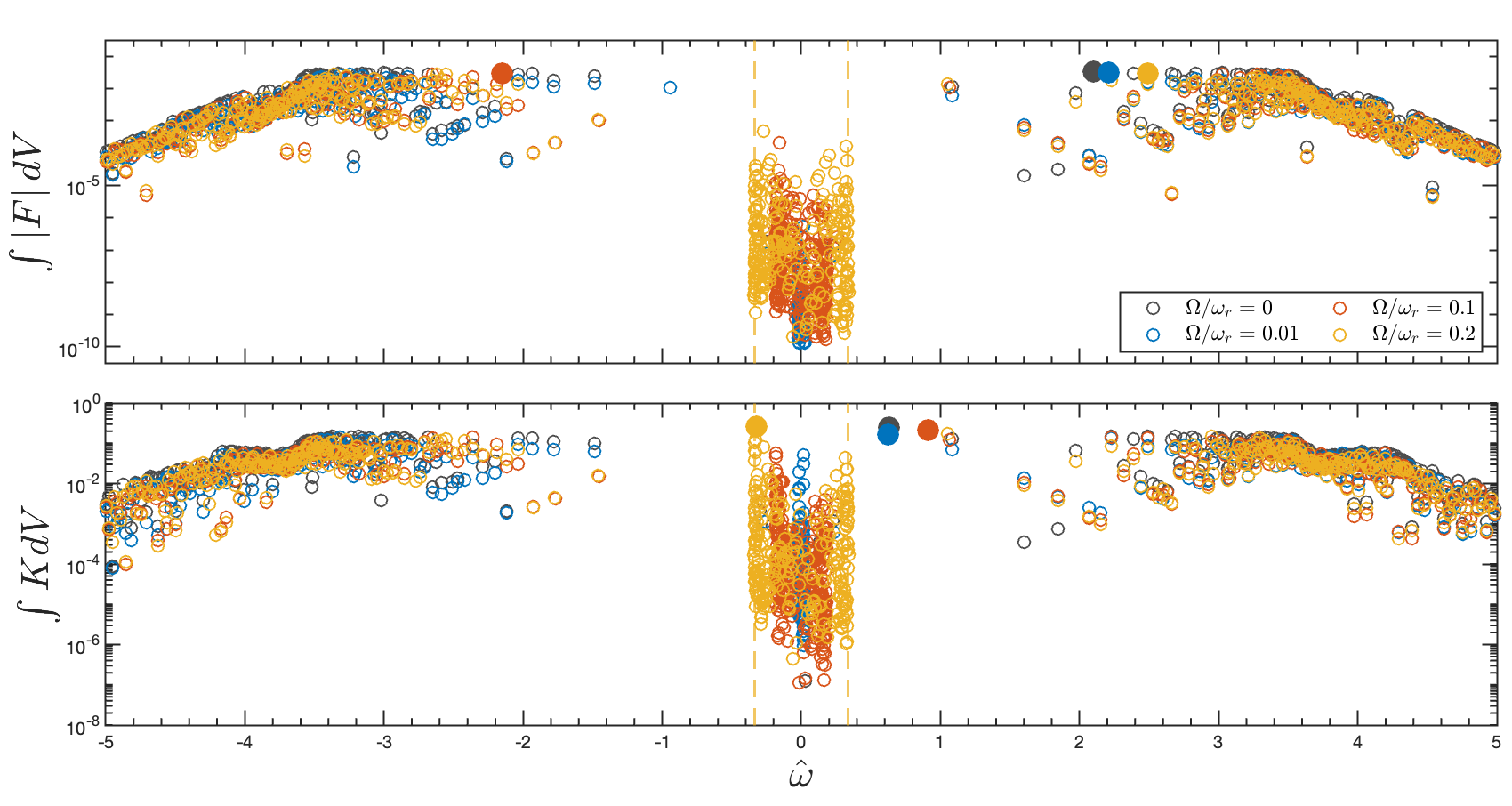}
    \caption{Model 1 with $m=2$. The symbols are the same as in Fig.\ref{fig:n1m0}.}
    \label{fig:n1m2}
\end{figure}

In addition to the axisymmetric mode ($m=0$), we also analyze the non-axisymmetric mode ($m=2$) as shown in Fig.\ref{fig:n1m2}. The magnitude of the radial energy flux and kinetic energy of the p modes increases significantly when $m=2$. For $m=0$, the perturbation is symmetric along the meridian, and the energy flux primarily propagates in the radial direction, possibly losing energy due to geometric attenuation. However, for $m=2$, the perturbation varies in the azimuthal direction, causing the energy flux to propagate along a spiral path, which may reduce geometric attenuation. Besides, the radial energy flux is dominated by p modes, while the kinetic energy is dominated by either p modes or r modes ($\Omega/\omega_{r}=0.2$). Notably, rotation has a stronger influence on non-axisymmetric modes. The wave frequency corresponding to $max\int\left|F\right|dV$ or $max\int KdV$ for $m=2$ varies with the rotation rate. 

\begin{figure}
	\centering
    \includegraphics[width=0.9\textwidth]{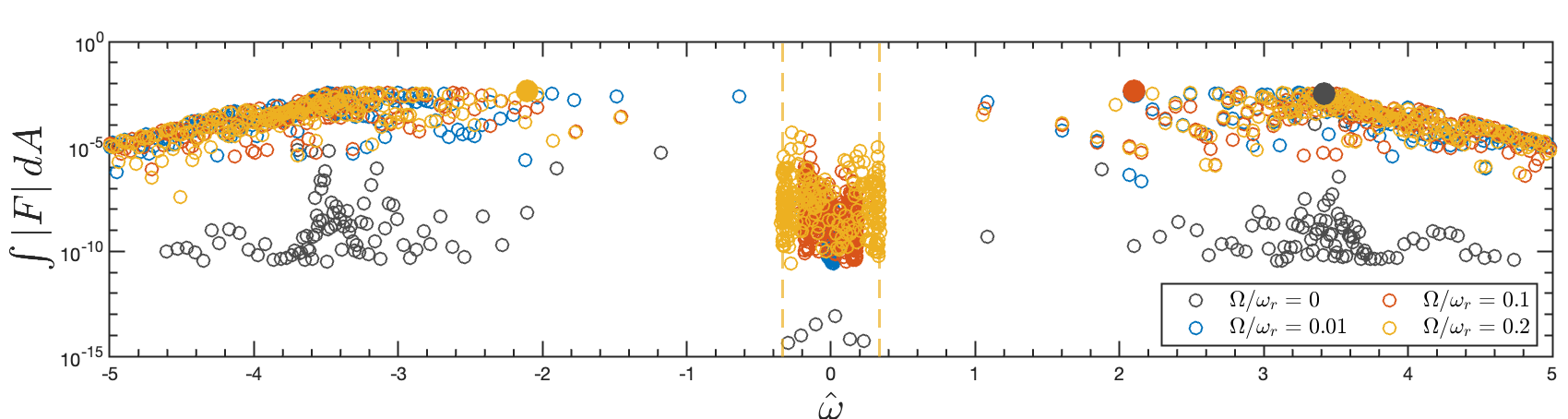}
	\caption{Model 1 with $m=2$ at $\phi=0$. Radial energy flux as a function of wave frequency $\hat{\omega}$ ($\hat{\omega}=sign(\omega)*\ln\left(1+\left|\omega/\omega_{r}\right|\right)$). Solid dots represent $\hat{\omega}$ corresponding to max$\int \left|F\right|dA$ ($dA=r^2\sin\theta drd\theta$). The other symbols are the same as in Fig.\ref{fig:n1m0}. 
 }\label{fig:n1m2s}
\end{figure}

The radial energy flux of the p modes do not have substantial change at different rotation rates, as shown in Fig.\ref{fig:n1m2}. However, when we calculate the energy flux in a certain meridional plane at $\phi=0$ as shown in Fig.\ref{fig:n1m2s}, the energy flux of the p modes increases by at least two to three orders of magnitude in the rotating case compared to the non-rotating case \citep{kippenhahn1990stellar,christensen2003lecture,goupil2009effects,aerts2010asteroseismology,schwarzschild2015structure}. Moreover, this increase in energy flux appears to be almost independent of the rotation rate. The fact that rotation facilitates wave propagation at certain meridional planes suggests that the internal dynamics of a rotating celestial body is radically different from a non-rotating body \citep{2006A&A...455..607L,2006A&A...455..621R,lovekin2008radial,reese2010oscillations,deupree2011theoretical,ballot2013numerical}. We give a tentative explanation. The Coriolis force modifies the spatial structure of p modes but it has little impact on their frequencies or propagation speeds \citep{chaboyer1995rotation}, so that the fundamental oscillation period of the p modes remains unchanged. However, the Coriolis force redistributes density perturbation. This results in local enhancements and reductions in the radial energy flux, but the overall radial energy flux remains almost unchanged. As comparison, kinetic energy which is not relevant to density perturbation does not significantly change. In a word, the Coriolis force indirectly affects the energy flux by changing the mass distribution.

\subsection{Rigid core model}
\begin{figure}
    \centering
      \includegraphics[width=0.9\textwidth]{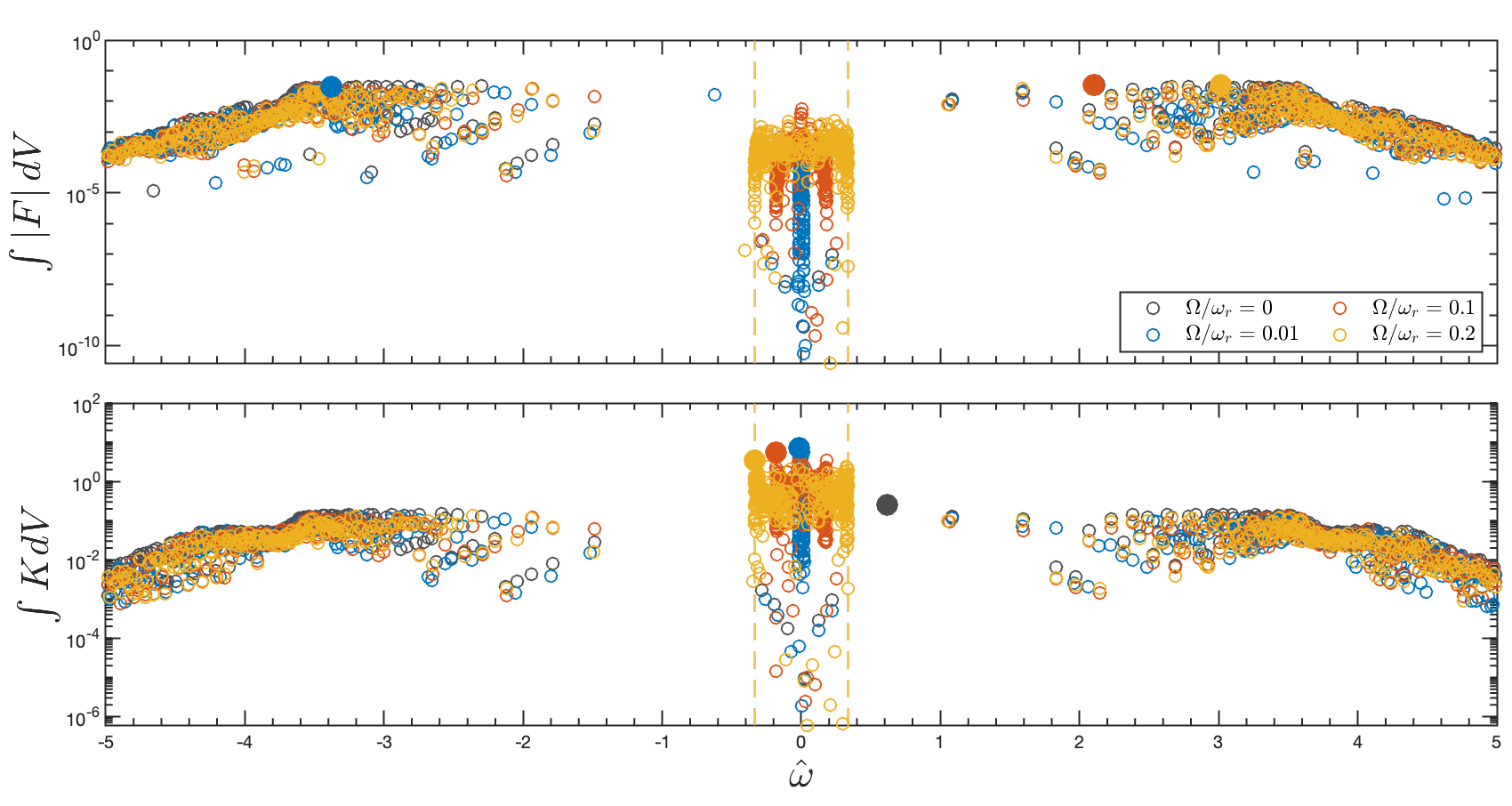}
    \caption{Model 2 with $m=2$. The symbols are the same as in Fig.\ref{fig:n1m0}. The gray and red solid dots nearly overlap, making them appear as a single point in the top panel.}
    \label{fig:n2m2}
\end{figure}

\begin{figure}
    \centering
      \includegraphics[width=0.9\textwidth]{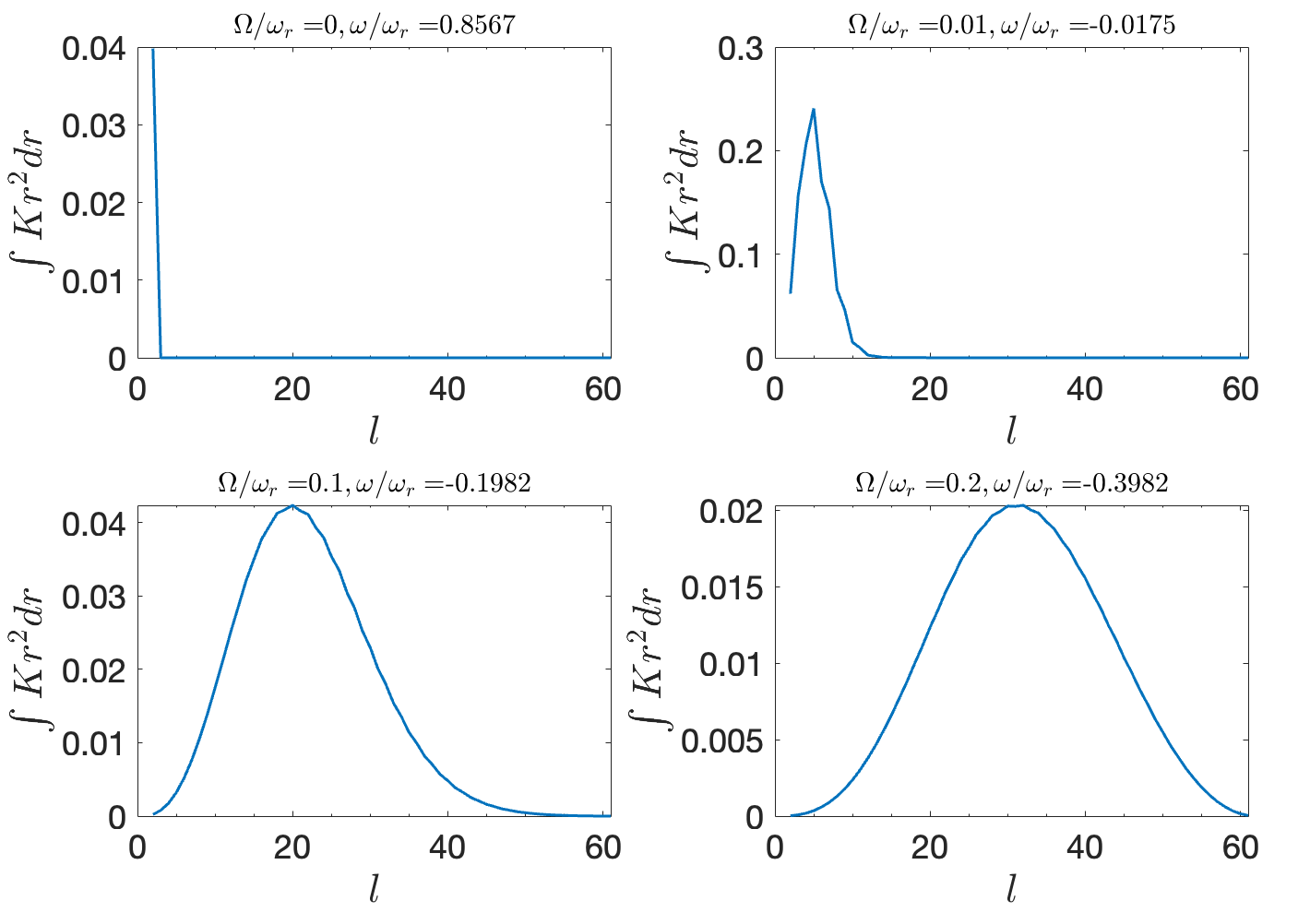}
    \caption{Model 2 with $m=2$. Kinetic energy as a function of degree $l$. For different rotation rates, We choose the mode that contains the largest kinetic energy in the isentropic fluid. }
    \label{fig:es}
\end{figure}

We now study Model 2: isentropic fluid surrounding a compact rigid core ($r\le0.1R$). Fig.\ref{fig:n2m2} shows the absolute value of radial energy flux $\left|F\right|$ and kinetic energy $K$ as a function of the wave frequency $\hat{\omega}$ with $m=2$. We can see that radial energy flux is dominated by p modes but kinetic energy by r modes under the influence of rotation. The difference between the axisymmetric and non-axisymmetric modes is similar to that in Model 1, i.e., rotation has a more striking influence on non-axisymmetric mode, so that we do not show the axisymmetric mode. Moreover, in the presence of a core, the radial energy flux and kinetic energy carried by r modes are improved compared to the full sphere model. It is known that inertial waves that bounce back and forth between the core and outer boundary can induce the attractors and the thin shear layers because of the wave emission at the singular critical latitude on the core \citep{stewartson1969pathological,rieutord2000,ogilvie2009}. Consequently, the influence of r modes on radial energy flux and kinetic energy is stronger with a compact core than without a compact core. The singularity of critical latitude is smoothed with numerical viscosity. Fig.\ref{fig:es} shows the resolution, i.e., kinetic energy as a function of $l$, and it is clear that our calculations are well resolved.

\subsection{Dilute core model} 
We move to the most interesting case, Model 3: isentropic fluid surrounding a dilute core. Recent Juno gravitational measurements suggest an extended dilute core in Jupiter, as opposed to a compact core \citep{2017GeoRL..44.4649W,2022PSJ.....3..185M,2024Icar..41115955M}. In this section, we consider a Jupiter model with a dilute core, treated as a stably stratified layer that supports g modes (gravity waves) restored by buoyancy, and g modes cannot exist in a convective envelope, i.e., their amplitude decays exponentially as they propagate from the stably stratified layer into the convective region. Unlike g modes, r modes can survive in both stably stratified layer and convective layers. The mixed gravito-inertial waves (GIWs) propagate under the influence of both buoyancy and Coriolis forces in a stably stratified layer \citep{1999JFM...398..271D,2017PhRvD..96h3005X,2020ApJ...903...90A}.  GIWs can propagate from the stably stratified layer into the convective layer, and vice versa. If such waves can be detected at Jupiter's surface, they would provide more information about Jupiter's internal structure.

\begin{figure}
    \centering
       \includegraphics[width=0.9\textwidth]{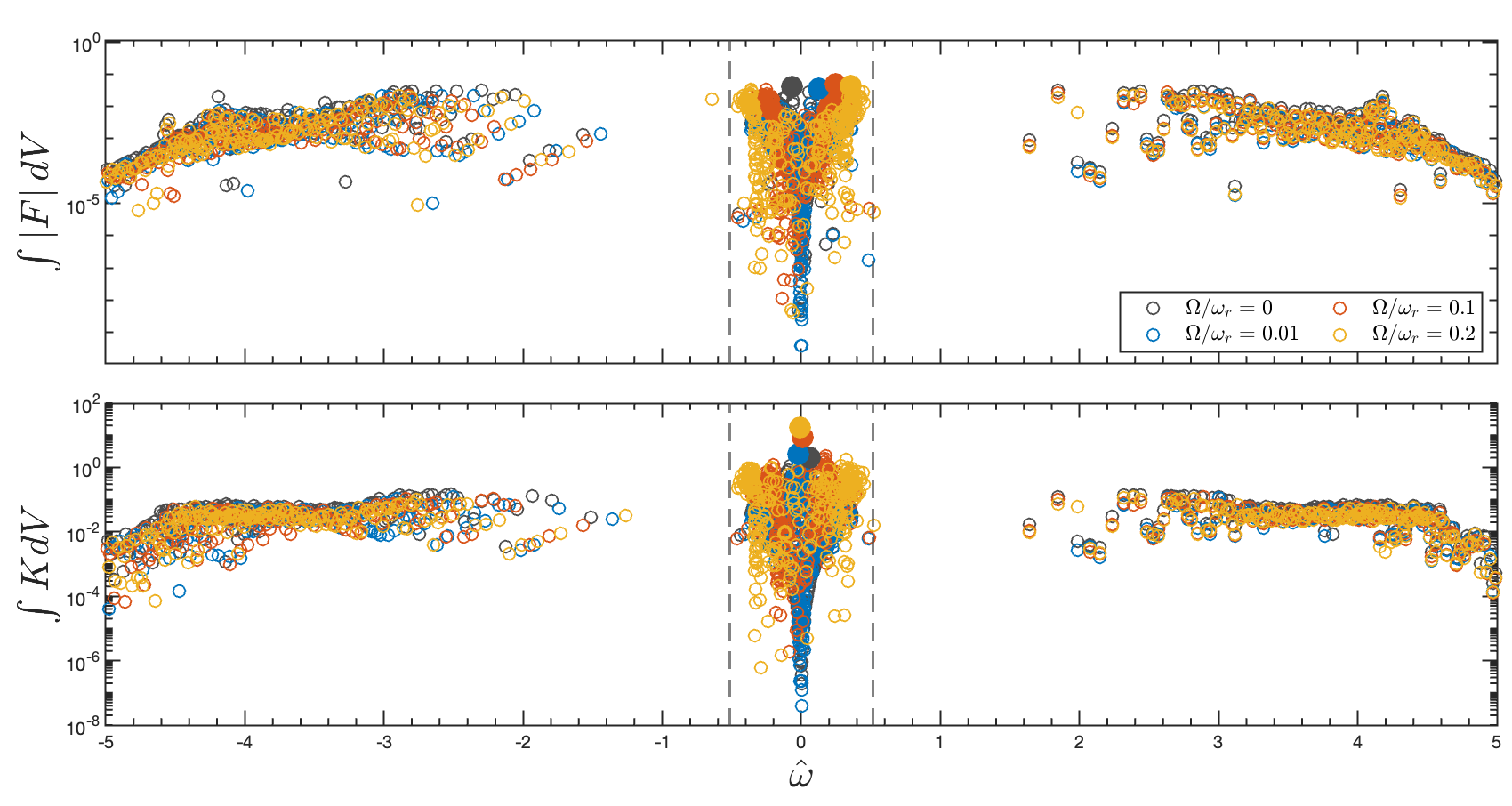}
    \caption{Model 3 with $m=2$. The symbols are the same as in Fig.\ref{fig:n1m0}. The frequency of GIWs (degenerate into pure g modes at $\Omega=0$) lies between the two gray dashed lines, while the region outside corresponds to p modes.}
    \label{fig:n3m2}
\end{figure}

Fig.\ref{fig:n3m2} shows radial energy flux and kinetic energy as a function of the wave frequency with $m=2$. The region enclosed by the two gray dashed lines represents the frequency range of GIWs, which degenerate into pure g modes at $\Omega/\omega_{r}=0$. Frequencies outside this range correspond to p modes. In the non-rotating case, compared with Fig.\ref{fig:n1m2} in Model 1, the radial energy flux and kinetic energy are both dominated by g modes. Moreover，the magnitude of the radial energy flux carried by some p modes also decreased. This suggests that the stable stratification suppresses the radial propagation of p modes to some extent. In the rotating case, the energy flux and kinetic energy are both dominated by GIWs with energy flux by fast GIWs (higher frequencies) and kinetic energy by slow GIWs (lower frequencies). The frequency of g modes is sufficiently high and comparable to the sound wave frequency, as shown in Fig.\ref{fig:bg2}, so that GIWs are capable to carry energy flux. 

\begin{figure}
	\centering
    \includegraphics[scale=0.3]{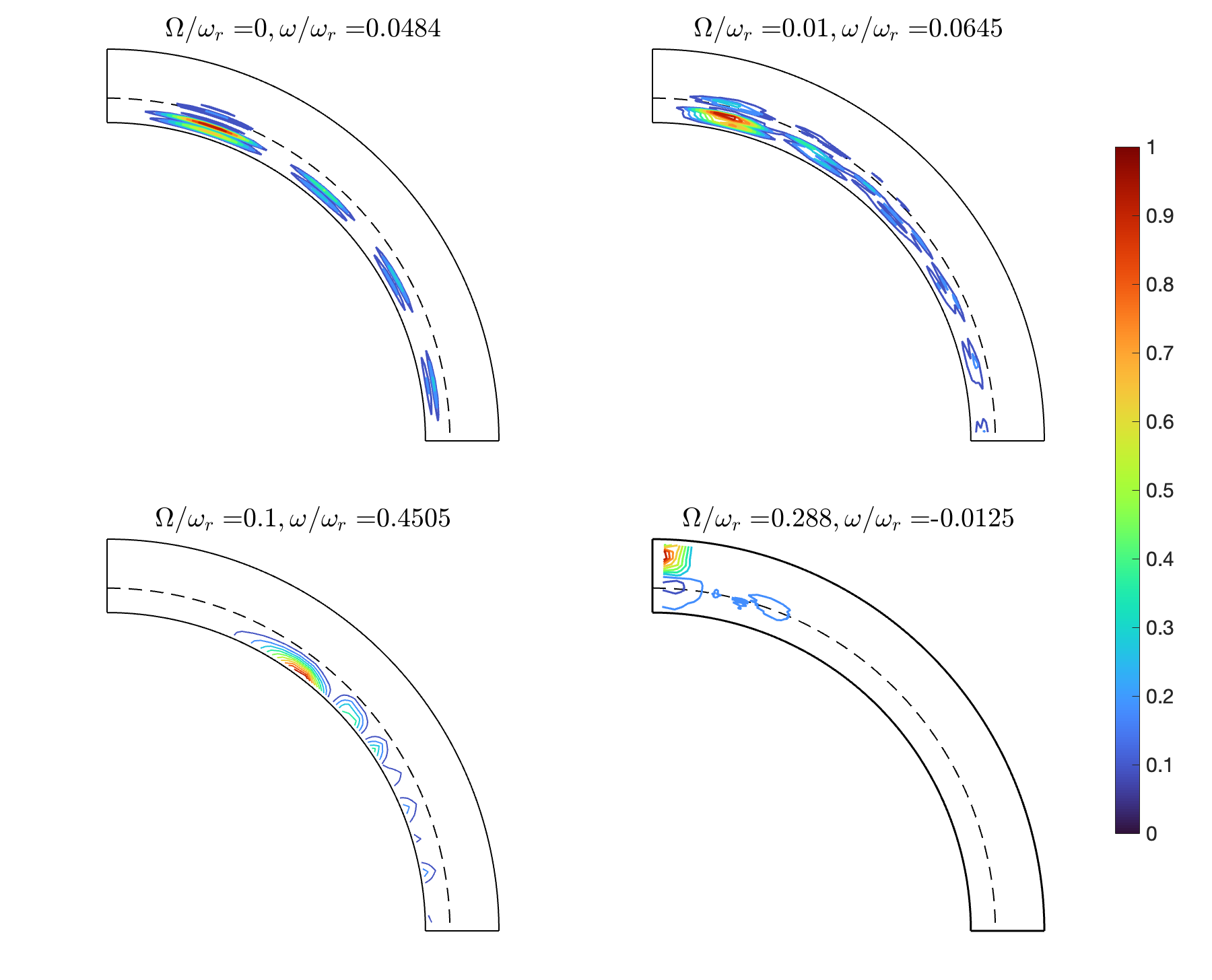}
	\caption{The absolute value of the radial energy flux $\left|F\right|$ at the transition layer between the dilute core and the convective envelope, with the radius ranging from 0.65$R$ to 0.8$R$ for $m=2$ and $\left|\omega/\omega_{r}\right|<0.5$, is shown. For different rotation rates, we choose the mode that contains the largest radial energy flux in the transition layer. The black dashed line marks the location of the interface between the dilute core and the convective envelope, at 0.7$R$. The bottom-right panel shows the wave transmission at Jupiter's rotation rate $\Omega=0.288$. The white areas indicate that the energy flux value is too small and negligible.
 }\label{fig:n1_tl}
\end{figure}

Then we investigate the wave transmission across the transition layer between the dilute core and convective envelope with respect to the non-axisymmetric mode ($m=2$). Fig.\ref{fig:n1_tl} shows the absolute value of radial energy flux at the transition layer between the dilute core and the convective envelope, with the radius ranging from 0.65$R$ to 0.8$R$. We select the mode in the range $|\omega/\omega_{r}|<0.5$ to avoid the fast p modes and containing $max\int\left|F\right|dV$ in transition layer. Thus, the internal wave transmission involves g modes, r modes, and GIWs. In order to compare the wave transmission at different rotation rates, we normalize the radial energy flux. The black dashed line marks the location of the interface 0.7$R$ between the dilute core and the convective envelope. In the absence of rotation, the energy flux is below the interface and the wave cannot transmit. This is exactly the characteristic of g modes, i.e., the group velocity of gravity waves is almost on the spherical surface. When rotation rate increases, the GIW gradually transmits across the interface. When rotation rate reaches its maximum 0.288 (Jupiter's dimensionless rotation rate), the GIW transmits at the high latitudes. That fast rotation facilitates the GIWs' transmission at high latitudes is because of the characteristic of r modes, i.e., the group velocity of inertial waves is almost along the rotational axis. 

Our result about wave transmission is qualitatively consistent with the analytical work by \citet{2020ApJ...890...20W,2020ApJ...899...88W}, i.e., rotation can enhance wave transmission. Seismology on Saturn's rings have revealed the stably stratified layer in the deep interior of Saturn \citep{2014AGUFM.P21E..03F,mankovich2019cassini,mankovich2021diffuse,mankovich2023saturn}. If Saturn has a stable stratification, then g modes can be excited in this region. Rotation provides an additional restoring force, allowing some of the g modes energy to leak out to the outer convective layers, thereby affecting the external gravitational field. This enhanced gravitational effect may be the source of the resonant g-modes observed in Saturn's rings. There are currently no similar observations for Jupiter, we may find Jupiter's internal structure by observing the oscillation signals in its polar region.

\section{CONCLUSIONS and discussions} \label{sec:conclusions}

In this paper, we study three models (full sphere, rigid core and dilute core) of Jupiter's internal structure with the fixed polytropic index $n=1$ but varying adiabatic index $\Gamma$. We calculate the adiabatic oscillation modes of compressible, self-gravitating and uniformly rotating polytropic model. Our analysis includes the Coriolis force but neglects the rotational distortion by centrifugal force. We focus on the different core structures on wave propagation and transmission in Jupiter's interior, that are relevant to the radial energy flux and kinetic energy of different modes. The findings of this study are summarized as follows.

In all three models, the radial energy flux is dominated by fast modes with higher frequencies, while the kinetic energy is carried primarily by slow modes with lower frequencies. This behavior is because the high-frequency modes with fast propagation efficiently carry energy away but the low-frequency modes with slow propagation retain energy. Rotation has a greater effect on non-axisymmetric modes (in addition to $m=2$ we calculated the other $m$'s and the results are similar). A rigid core facilitates the propagation of r modes.  Moreover, rotation enhances the transmission of GIWs at high latitudes because of the direction of group velocity of inertial waves. Therefore, if we observe the oscillation signals of GIWs in polar region of a rapidly rotating giant planet, we may find its internal structures.

There are some caveats, which should be considered in future. Firstly, we neglect the centrifugal force for simplicity, and the future study can take into account the spheroidal geometry. Secondly, planetary interior exhibits differential rotation and meridional circulation which can significantly affect the oscillation modes. Thirdly, the magnetic fields, which induce Alfv\'en waves, can couple with pressure, rotation and stratification to influence the oscillation modes and consequently the propagation and transmission of internal waves. Fourthly，Juno data can infer the existence of a flat metal-rich interior and a metal-poor outer region, indicating a dilute core \citep{2022PSJ.....3..185M,2024Icar..41115955M}. Physically, such a core would be convectively mixed on a timescale of a few hundred million years \citep{2024ApJ...977..227K,2025ApJ...979..243A}. Therefore, to better understand Jupiter's internal structure and evolution, it will be very helpful to study the wave propagation and transmission in the background of heavy metal gradient. Finally, a more realistic EOS may deviate from the $n=1$ polytrope when accounting for the gas-metallic fluid phase transition \citep{2013ApJ...774..148M,2019ApJ...872...51C} or heavy elements in the diluted core. This effect may alter the density distribution, thereby affecting quantities such as the buoyancy frequency, which in turn influence wave propagation and transmission. It is necessary to adopt a more realistic EOS  in future work. For example, \citet{2021PSJ.....2..241N} used a polytrope EOS but modified the density profile accordingly when modeling the interiors of Jupiter and Saturn.

\section*{Acknowledgment}
Yuru Xu thanks Yufeng Lin, Qiang Hou and Zehao Su for their helpful discussions. Xing Wei is financially supported by NSFC (12041301).

\appendix
\section{Energy balance} \label{appendix}
The equations of mass and momentum of inviscid fluid read
\begin{equation} \label{eq:mass_eb}
	\frac{\partial\rho}{\partial t}+\boldsymbol{\nabla}\cdot\left(\rho\boldsymbol{u}\right)=0,
\end{equation}
\begin{equation} \label{eq:mom_eb}
     \rho\left(\frac{\partial}{\partial t}+\boldsymbol{u}\cdot\boldsymbol{\nabla}\right)\boldsymbol{u}=-\boldsymbol{\nabla}P+2\rho\boldsymbol{u \times\Omega}-\rho\boldsymbol{\nabla}\Phi.    
\end{equation}
Taking the dot product of Eq.(\ref{eq:mom_eb}) with $\boldsymbol{u}$, by virtue of mass conservation Eq.(\ref{eq:mass_eb}), we can derive the equation of kinetic and gravitational energies
\begin{equation}  \label{eq:td3}
	\frac{\partial}{\partial t}\left(\frac{1}{2}\rho u^{2}+\rho \Phi\right)+\boldsymbol{\nabla}\cdot\left[\boldsymbol{u}\left(\frac{1}{2}\rho u^{2}+P+\rho\Phi\right)\right]=P\boldsymbol{\nabla}\cdot\boldsymbol{u}+\rho\frac{\partial \Phi}{\partial t}.
\end{equation}
Isentropic layer is adiabatic and stably stratified layer is in thermal balance (heat influx is equal to cooling flux at all radii) such that in both layers there is no net heat exchange, that is exactly what the polytropic model suggests. Therefore the equation of internal energy reads
\begin{equation} \label{eq:thermodynamic}
	\frac{de}{dt}+P\frac{d}{dt}\left(\frac{1}{\rho}\right)=0.
\end{equation}
where $e$ is the internal energy per unit mass.\\ 
Substituting $e=nP/\rho$ into Eq.(\ref{eq:thermodynamic}), by virtue of mass conservation Eq.(\ref{eq:mass_eb}), we can derive the equation of internal energy per unit volume
\begin{equation}  \label{eq:td2}
	\frac{\partial \left(nP\right)}{\partial t}+\boldsymbol{\nabla}\cdot\left(nP\boldsymbol{u}\right)=-P\boldsymbol{\nabla}\cdot\boldsymbol{u}.
\end{equation}
Adding (\ref{eq:td3}) and Eqs.(\ref{eq:td2}) leads to the total energy equation
\begin{equation} \label{eq:tee}
	\frac{\partial}{\partial t}\left(\frac{1}{2}\rho u^{2}+nP+\rho\Phi \right)+\boldsymbol{\nabla}\cdot\left[\boldsymbol{u}\left(\frac{1}{2}\rho u^{2}+(1+n)P+\rho\Phi\right)\right]=\rho\frac{\partial \Phi}{\partial t}.
\end{equation}
The second term on LHS in the divergence represents the energy flux. We perturb the energy flux, take the time average such that the first-order terms vanish, and neglect the orders higher than second-order,
\begin{equation} \label{eq:lptee}
	\frac{\partial}{\partial t}\left(\frac{1}{2}\rho_{0}u^{\prime 2}+\rho^{\prime}\Phi^{\prime}\right)+\boldsymbol{\nabla}\cdot \left[\left(1+n\right)p^{\prime}\boldsymbol{u}^{\prime}+\Phi_{0}\rho^{\prime}\boldsymbol{u}^{\prime}+\rho_{0}\Phi^{\prime}\boldsymbol{u}^{\prime}\right]=\rho^{\prime}\frac{\partial \Phi^{\prime}}{\partial t}.
\end{equation}
In the energy flux, we neglect the term associated with gravitational potential $\Phi'$, i.e., Cowling approximation which works well for short waves (we have validated this approximation in our calculations), to find the radial energy flux
\begin{equation}
F=\left(1+n\right)p^{\prime}{u_r}^{\prime}+\Phi_{0}\rho^{\prime}{u_r}^{\prime}.
\end{equation}

\bibliography{paper}{}
\bibliographystyle{aasjournal}

\end{CJK*}
\end{document}